\pdfoutput=1
\documentclass{llncs}

\usepackage{lineno}

\usepackage{amsfonts,amsmath}

\usepackage{graphicx}
\usepackage{url}
\usepackage{verbatim}
\usepackage{color}
\definecolor{lgray}{gray}{0.95}
\definecolor{lblue}{rgb}{0.90,0.90,1.00}
\definecolor{lyellow}{rgb}{1.00,1.00,0.70}

\usepackage{listings}
\newtheorem{ex}{Example}

\newenvironment{codex}{\small\verbatim}{\endverbatim\normalsize}

\lstnewenvironment{code}
    {\lstset{}%
      \csname lst@SetFirstLabel\endcsname}
    {\csname lst@SaveFirstLabel\endcsname}
\newcommand{\BI}[0]{\begin{itemize}}
\newcommand{\EI}[0]{\end{itemize}}

\newcommand{\BE}[0]{\begin{enumerate}}
\newcommand{\EE}[0]{\end{enumerate}}

\newcommand{\BX}[0]{\begin{ex}}
\newcommand{\EX}[0]{\end{ex}}
\newcommand{\BV}[0]{\begin{verbatim}}
\newcommand{\EV}[0]{\end{verbatim}}

\def \bscale1 {0.25}
\def \bscale {0.25}
\def \N {\mathbb{N}}

\begin{document}

\title{ 
A Hiking Trip Through the Orders of Magnitude: Deriving Efficient 
Generators for Closed Simply-Typed Lambda Terms and Normal Forms
}

\author{Paul Tarau}
\institute{
   {Department of Computer Science and Engineering}\\
   {University of North Texas}\\ 
   {\em tarau@cse.unt.edu}\\
}

\maketitle

\begin{abstract}
Contrary to several other families of lambda terms,
no closed formula or generating function 
is known and none of the sophisticated 
techniques devised in analytic combinatorics
can currently help with counting or generating
the set of {\em simply-typed closed lambda terms} 
of a given size.

Moreover,  their asymptotic scarcity among
the set of closed lambda terms makes counting them
via brute force generation and type 
inference quickly intractable, with previous published
work  showing counts for them only up to size 10.

By taking advantage of the synergy between logic variables,
unification with occurs check and efficient backtracking 
in today's Prolog systems, we climb 4 orders
of magnitude above previously known 
counts by deriving 
progressively faster Horn Clause programs 
that generate and/or count the set of closed simply-typed 
lambda terms of sizes up to 14.
A similar count for {\em closed simply-typed normal forms}
is also derived up to size 14.


\end{abstract}

\section{Introduction}

Generation of lambda terms \cite{bar84} has practical
applications to testing compilers that rely on
lambda calculus as an intermediate language,
as well as in generation of random tests
for user-level programs and data types. 
At the same time, several instances of 
lambda calculus are of significant theoretical
interest given their correspondence with
logic and proofs.

Simply-typed lambda terms \cite{hindley2008lambda,bar93} enjoy
a number of nice properties, among which strong normalization
(termination for all evaluation-orders),
a cartesian closed category mapping and a set-theoretical
semantics. More importantly, via the Curry-Howard correspondence
lambda terms that are {\em inhabitants} of simple types
can be seen as proofs for tautologies in {\em minimal logic}
which, in turn, correspond to the types.
Extended with a fix-point operator, simply-typed lambda terms
can be used as the intermediate language for compiling
Turing-complete functional languages.
Random generation of simply-typed lambda terms can also help with
automation of debugging compilers for functional programming
languages  \cite{palka11}.
 
Recent work on the combinatorics of
lambda terms \cite{grygielGen,ranlamb09,bodini11,normalizing13},
relying on recursion equations, generating functions
and techniques from analytic combinatorics \cite{flajolet09}
has provided counts for several families
of lambda terms and clarified important
properties like their asymptotic density.
With the techniques provided
by generating functions \cite{flajolet09},
it was possible to separate the
{\em counting} of the terms of a given size for
several families of lambda terms from
their more computation intensive  {\em generation}, resulting
in several additions (e.g., A220894, A224345, A114851) to
The On-Line Encyclopedia of Integer Sequences, \cite{intseq}.

On the other hand, the combinatorics
of simply-typed lambda terms, given the absence
of closed formulas, recurrence equations
or grammar-based generators, due to the intricate interaction
between type inference and the applicative structure
of lambda terms, has left important
problems open, including the very basic one of
counting the number of closed simply-typed
lambda terms of a given size.
At this point, obtaining counts for simply-typed
lambda terms requires going through the more computation-intensive 
generation process.

As a fortunate synergy, Prolog's sound unification of logic variables, 
backtracking and  definite clause grammars  have been shown to
provide compact combinatorial generation algorithms for
various families of lambda terms \cite{padl15,cicm15,ppdp15tarau,iclp15}.

For the case of simply-typed lambda terms, we have pushed (in the unpublished draft
\cite{arxiv_play15}) the counts in sequence A220471 of \cite{intseq} to cover
sizes {\tt 11} and {\tt 12}, each requiring  about one magnitude of extra
computation effort, simply by writing the generators in Prolog.
In this paper we focus on going two more magnitudes higher,
while also integrating
the results described in \cite{arxiv_play15}.
Using similar techniques, we achieve the same, for the special case
of simply-typed normal forms.

The paper is organized as follows.
Section \ref{lgen} describes our representation of lambda terms
and derives a generator for closed lambda terms.
Section \ref{ltypes} defines generators for well-formed type formulas.
Section \ref{lmerge} introduces a type inference algorithm
and then derives, step by step, efficient generators for simply-typed
lambda terms and simple types inhabited by terms of a given size.
Section \ref{lnormal} defines generators for 
closed lambda terms in normal form and then replicates
the derivation of an efficient generator for simply-typed
closed normal forms.
Section \ref{perf} aggregates our experimental 
performance data and section  \ref{disc} discusses 
possible extensions and future improvements.
Section \ref{rels} overviews related work and section
\ref{concl} concludes the paper.

The paper is structured as a literate Prolog program.
The  code has been tested with SWI-Prolog 7.3.8 and YAP 6.3.4.
It is also available as a separate file at
\url{http://www.cse.unt.edu/~tarau/research/2016/lgen.pro}.

\begin{codeh}
:-ensure_loaded(library(lists)).
\end{codeh}

\section{Deriving a generator for lambda terms} \label{lgen}

Lambda terms can be seen as  Motzkin trees \cite{StanleyEC}, also
called unary-binary trees, labeled with lambda binders at
their unary nodes and corresponding variables at the leaves.
We will thus derive a generator for them from a generator for
Motzkin trees.

\subsection{A canonical representation with logic variables}
We can represent lambda terms \cite{bar84} in Prolog using the constructors
{\tt a/2} for applications, {\tt l/2} for lambda abstractions
and {\tt v/1} for variable occurrences.
Variables bound by the lambdas and their occurrences are
represented as {\em logic variables}.
As an example, the lambda 
term $~\lambda a.(\lambda b.(a~(b~b)) ~ \lambda c.(a~(c~c)) )$
will be represented as 
{\tt l(A,a(l(B,a(v(A),a(v(B),v(B)))),l(C,a(v(A),a(v(C),v(C))))))}.
As variables share a unique scope (the clause
containing them), this representation assumes that 
{\em distinct variables are used
for distinct scopes induced by the lambda binders} in 
terms occurring in a given Prolog clause.

Lambda terms might contain {\em free variables} not associated to any binders.
Such terms are called {\em open}. A {\em closed} term is such that each variable
occurrence is associated to a binder.


\subsection{Generating Motzkin trees}
Motzkin-trees (also called binary-unary trees)
have internal nodes of arities 1 or 2. Thus they can be seen
as a skeleton of lambda terms that ignores binders and 
variables and their leaves.

The predicate {\tt motzkin/2} generates Motzkin trees
with {\tt S} internal and leaf nodes. 
\begin{code}
motzkin(S,X):-motzkin(X,S,0).

motzkin(v)-->[].
motzkin(l(X))-->down,motzkin(X).
motzkin(a(X,Y))-->down,motzkin(X),motzkin(Y).

down(s(X),X).
\end{code}

Motzkin-trees, with leaves assumed of size 1 are counted by the sequence
{\tt A001006} in \cite{intseq}. Alternatively, as in our case, when leaves are
assumed of size 0, we obtain binary-unary trees with {\tt S}
internal nodes, counted by the entry {\tt A006318} (Large Schr\"oder Numbers)
of \cite{intseq}.

Note  the use of the predicate {\tt down/2},  that assumes natural numbers
in {\em unary notation}, 
with $n$ {\tt s/1} symbols wrapped around {\tt 0} to
denote $n \in \N$. As our combinatorial generation algorithms
will usually be tractable for values of $n$ below {\tt 15}, the use
of unary notation is comparable (and often  slightly faster)
than the call to arithmetic built-ins. Note also that this
leads, after the DCG translation, to ``pure'' Prolog programs
made exclusively of Horn Clauses, as the DCG notation
can be eliminated by threading two extra arguments controlling
the size of the terms.

To more conveniently call these generators with 
the usual natural numbers
we  define the converter {\tt n2s} as follows.
\begin{code}
n2s(0,0).
n2s(N,s(X)):-N>0,N1 is N-1,n2s(N1,X).
\end{code}

\BX
Motzkin trees with 2 internal nodes.
\begin{codex}
?- n2s(1,S),motzkin(S,T).
S = s(0), T = l(v) ;
S = s(0), T = a(v, v) .
\end{codex}
\EX

\subsection{Generating closed lambda terms} \label{lambda}
We derive a generator for closed lambda terms by
adding logic variables as labels to their
binder and variable nodes, while ensuring
that the terms are closed, i.e.,
that the function mapping variables to
their binders is total.

The predicate {\tt lambda/2} 
builds a list of logic variables as it generates binders. 
When generating a leaf variable,
it picks ``nondeterministically'' one of the binders among
the list of binders available, {\tt Vs}. As in the case of Motzkin trees, the predicate
{\tt down/2} controls the number of internal nodes.

\begin{code}
lambda(S,X):-lambda(X,[],S,0).

lambda(v(V),Vs)-->{member(V,Vs)}.
lambda(l(V,X),Vs)-->down,lambda(X,[V|Vs]).
lambda(a(X,Y),Vs)-->down,lambda(X,Vs),lambda(Y,Vs).
\end{code}

The sequence {\tt A220471} 
in \cite{intseq}
contains counts for lambda terms of increasing sizes,
with {\em size defined as the number of internal nodes}.

\BX
Closed lambda terms with 2 internal nodes.
\begin{codex}
?- lambda(s(s(0)),Term).
Term = l(A, l(B, v(B))) ;
Term = l(A, l(B, v(A))) ;
Term = l(A, a(v(A), v(A))) .
\end{codex}
\EX

\section{A visit to the other side: the language of types} \label{ltypes}

As a result of the Curry-Howard correspondence,
the language of types is isomorphic with that
of {\em minimal logic}, with binary trees having variables
at leaf positions and the implication operator (``\verb~->~'')
at internal nodes. We will rely on the right associativity
of this operator in Prolog, that matches the standard
notation in type theory.

The predicate {\tt type\_skel/3} generates all binary trees
with given number of internal nodes and labels their leaves
with unique logic variables. It also collects the variables
to a list returned as its third argument.
\begin{code}
type_skel(S,T,Vs):-type_skel(T,Vs,[],S,0).

type_skel(V,[V|Vs],Vs)-->[].
type_skel((X->Y),Vs1,Vs3)-->down,
  type_skel(X,Vs1,Vs2),
  type_skel(Y,Vs2,Vs3).
\end{code}
Type skeletons are counted by the Catalan numbers
(sequence {\tt A000108} in \cite{intseq}).
\BX
All type skeletons for N=3.
\begin{codex}
?- type_skel(s(s(s(0))),T,_).
T =  (A->B->C->D) ;
T =  (A-> (B->C)->D) ;
T =  ((A->B)->C->D) ;
T =  ((A->B->C)->D) ;
T =  (((A->B)->C)->D) .
\end{codex}
\EX

The next step  toward generating the set of all type formulas
is observing that logic variables define equivalence
classes that can be used to generate partitions
of the set of variables, simply by
selectively unifying them.

The predicate {\tt mpart\_of/2}
takes a list of distinct logic variables
and generates partitions-as-equivalence-relations
by unifying them ``nondeterministically''.
It also collects the unique variables, defining
the equivalence classes as a list given by its second argument.
\begin{code}
mpart_of([],[]).
mpart_of([U|Xs],[U|Us]):-
  mcomplement_of(U,Xs,Rs),
  mpart_of(Rs,Us).
\end{code}

To implement a set-partition generator,
we will split a set repeatedly in subset+complement
pairs with help from the predicate {\tt mcomplement\_of/2}.
\begin{code}
mcomplement_of(_,[],[]).
mcomplement_of(U,[X|Xs],NewZs):-
  mcomplement_of(U,Xs,Zs),
  mplace_element(U,X,Zs,NewZs).

mplace_element(U,U,Zs,Zs).
mplace_element(_,X,Zs,[X|Zs]).
\end{code}
To generate set partitions of a set of variables
of a given size, we build a list
of fresh variables with the equivalent of
Prolog's {\tt length} predicate working
in unary notation, {\tt len/2}.
\begin{code}
partitions(S,Ps):-len(Ps,S),mpart_of(Ps,_).

len([],0).
len([_|Vs],s(L)):-len(Vs,L).
\end{code}

The count of the resulting set-partitions (Bell numbers)
corresponds to the entry {\tt A000110} 
in \cite{intseq}.

\BX
Set partitions of size 3 expressed as variable equalities.
\begin{codex}
?- partitions(s(s(s(0))),P).
P = [A, A, A] ;
P = [A, B, A] ;
P = [A, A, B] ;
P = [A, B, B] ;
P = [A, B, C].
\end{codex}
\EX

We can then define the language of formulas in minimal
logic, among which tautologies will correspond to
simple types, as being generated by the predicate {\tt maybe\_type/3}.
\begin{code}
maybe_type(L,T,Us):-type_skel(L,T,Vs),mpart_of(Vs,Us).
\end{code}

\BX
Well-formed  formulas of minimal logic (possibly types) of size 2.

\begin{codex}
?- maybe_type(s(s(0)),T,_).
T =  (A->A->A) ;
T =  (A->B->A) ;
T =  (A->A->B) ;
T =  (A->B->B) ;
T =  (A->B->C) ;
T =  ((A->A)->A) ;
T =  ((A->B)->A) ;
T =  ((A->A)->B) ;
T =  ((A->B)->B) ;
T =  ((A->B)->C) .
\end{codex}
\EX
The 
sequence {\tt 2,10,75,728,8526,115764,1776060,30240210} 
counting these formulas corresponds to
the product of Catalan and Bell numbers.

\section{Merging the two worlds: generating simply-typable lambda terms} \label{lmerge}

One can observe that per-size counts of both the sets of
lambda terms and their potential types are very fast growing.
There is an important difference, though, between
computing the type of a given lambda term (if it exists)
and computing an inhabitant of a type (if it exists).
The first operation, called {\em type inference} is an efficient
operation (linear in practice) while the
second operation, called {\em the inhabitation problem}
is {\tt P-space} complete \cite{statman79}.

This brings us to design a type inference algorithm
that takes advantage of operations on logic variables.

\subsection{A type inference algorithm}

While in a functional language inferring types
requires implementing unification with occurs-check, 
as shown for instance in \cite{grygielGen}, 
this operation is available in Prolog as a built-in
predicate, optimized, for instance, in SWI-Prolog
\cite{swi}, to proceed incrementally,
only checking that no new cycles are introduced
during the unification step as such.

The predicate {\tt infer\_type/3} works by
using logic variables as dictionaries associating
terms to their types. Each logic variable is then bound
to a term of the form {\tt X:T} where {\tt X} will be a component
of a fresh copy of the term and {\tt T} will be its type.
Note that we create this new term
as the original term's variables end up
loaded with chunks of the partial types created
during the type inference process.

As logic variable bindings propagate between binders
and occurrences, this ensures that types are 
consistently inferred.

\begin{code}
infer_type((v(XT)),v(X),T):-unify_with_occurs_check(XT,X:T).
infer_type(l((X:TX),A),l(X,NewA),(TX->TA)):-infer_type(A,NewA,TA).
infer_type(a(A,B),a(X,Y),TY):-infer_type(A,X,(TX->TY)),infer_type(B,Y,TX).
\end{code}

\BX
illustrates typability of the term corresponding to the
{\tt S} combinator\\ $ \lambda x_0. \lambda x_1. \lambda x_2.((x_0~x_2)~(x_1~x_2))$\\
and untypabilty of the term corresponding to the {\tt Y} combinator\\ 
$ \lambda x_0.(\lambda x_1.(x_0~(x_1~x_1)) ~ ~\lambda x_2.(x_0~(x_2~x_2)) ) $.
\begin{codex}
?- infer_type(l(A,l(B,l(C,a(a(v(A),v(C)),a(v(B),v(C)))))),X,T),
   portray_clause((T:-X)),fail.
(A->B->C)-> (A->B)->A->C :-
	   l(D,l(F,l(E, a(a(v(D), v(E)), a(v(F), v(E)))))).
	    
?- infer_type(
     l(A,a(l(B,a(v(A),a(v(B),v(B)))),l(C,a(v(A),a(v(C),v(C)))))), X, T).
false.	    
\end{codex}
\EX

By combining generation of lambda terms with type inference we have
our first cut to an already surprisingly fast generator
for simply-typable lambda terms, able to generate in a few
hours counts for sizes {\tt 11} and {\tt 12} for
sequence {\tt A220471} in \cite{intseq}.

\begin{code}
lamb_with_type(S,X,T):-lambda(S,XT),infer_type(XT,X,T).
\end{code}

\BX
Lambda terms of size up to 3 and their types.
\begin{codex}
?- lamb_with_type(s(s(s(0))),Term,Type).
Term = l(A, l(B, l(C, v(C)))), Type =  (D->E->F->F) ;
Term = l(A, l(B, l(C, v(B)))), Type =  (D->E->F->E) ;
Term = l(A, l(B, l(C, v(A)))), Type =  (D->E->F->D) ;
Term = l(A, l(B, a(v(B), v(A)))), Type =  (C-> (C->D)->D) ;
Term = l(A, l(B, a(v(A), v(B)))), Type =  ((C->D)->C->D) ;
Term = l(A, a(v(A), l(B, v(B)))), Type =  (((C->C)->D)->D) ;
Term = l(A, a(l(B, v(B)), v(A))), Type =  (C->C) ;
Term = l(A, a(l(B, v(A)), v(A))), Type =  (C->C) ;
Term = a(l(A, v(A)), l(B, v(B))), Type =  (C->C).
\end{codex}
\EX

Note that, for instance, when one wants to select 
only terms having a given type,
this is quite inefficient. Next, we will show how to
combine size-bound term generation,
testing for closed terms and type inference into a single
predicate. This will enable more efficient querying
for terms inhabiting a given type, as one
would expect from Prolog's multi-directional
execution model, and more importantly for our purposes,
to climb two orders of magnitude higher
for counting simply-typed 
terms of size {\tt 13} and {\tt 14}.

\subsection{Combining term generation and type inference}

We need two changes to {\tt infer\_type}
to turn it into an efficient generator for simply-typed lambda terms.
First, we need to add an argument to control the size
of the terms and ensure 
termination, by calling {\tt down/2} for internal nodes.
Second, we need to generate the mapping between binders
and variables. We ensure this by borrowing the
{\tt member/2}-based mechanism used in the predicate
{\tt lambda/4} generating closed lambda terms in subsection \ref{lambda}.

The predicate {\tt typed\_lambda/3} does just that,
with helper from DCG-expanded {\tt typed\_lambda/5}.
\begin{code}
typed_lambda(S,X,T):-typed_lambda(_XT,X,T,[],S,0).

typed_lambda(v(V:T),v(V),T,Vs)--> {
   member(V:T0,Vs),
   unify_with_occurs_check(T0,T)
  }.
typed_lambda(l(X:TX,A),l(X,NewA),(TX->TY),Vs)-->down,
  typed_lambda(A,NewA,TY,[X:TX|Vs]).   
typed_lambda(a(A,B),a(NewA,NewB),TY,Vs)-->down,
  typed_lambda(A,NewA,(TX->TY),Vs),
  typed_lambda(B,NewB,TX,Vs).
\end{code}
Like {\tt lambda/4}, the predicate {\tt typed\_lambda/5}
relies on Prolog's DCG notation
to thread together the steps
controlled by the predicate {\tt down}.
Note also the nondeterministic use of
the built-in {\tt member/2} that enumerates
values for {\tt variable:type} pairs
ranging over the list 
of available pairs {\tt Vs},
as well as the use of {\tt
unify\_with\_occurs\_check}
to ensure that unification
of candidate types 
does not create cycles.

\BX
A term of size 15 and its type.
\begin{codex}
l(A,l(B,l(C,l(D,l(E,l(F,l(G,l(H,l(I,l(J,l(K,
           a(v(I),l(L,a(a(v(E),v(J)),v(J)))))))))))))))
M->N->O->P-> (Q->Q->R)->S->T->U-> ((V->R)->W)->Q->X->W
\end{codex}
\EX

We will discuss exact performance data later, but let's note
here that this operation
brings down by an order of magnitude
the computational effort to generate simply-typed terms.
As expected, the number of solutions is computed as the sequence 
{\tt A220471} in \cite{intseq}.
Interestingly, by {\em interleaving} generation of closed terms and
type inference in the predicate {\tt typed\_lambda}, the
time to generate all the closed simply-typed terms
is actually shorter than the time to generate
all closed terms of the same size, e.g., 17.123 vs.  31.442 seconds for size 10 
with SWI-Prolog. As, via the Curry-Howard isomorphism, closed simply typed terms
correspond to proofs of tautologies in minimal logic, co-generation of terms and
types corresponds to co-generation of tautologies and their proofs for
proofs of given length.

\subsection{One more trim: generating inhabited types}

Let's first observe that the actual lambda
term does not need to be built, provided
that we mimic exactly the type inference operations
that one would need to perform to ensure it
is simply-typed. It is thus safe to remove
the first argument of
{\tt typed\_lambda/5} as well
as the building of the fresh copy performed
in the second argument.
To further simplify the code, we can also make
the DCG-processing of the size computations explicit,
in the last two arguments.

This gives the predicate {\tt inhabited\_type/4}
and then {\tt inhabited\_type/2},
that generates {\em all types having inhabitants of
a given size}, but omits the inhabitants as such.
\begin{code}
inhabited_type(X,Vs,N,N):-
  member(V,Vs),
  unify_with_occurs_check(X,V).
inhabited_type((X->Xs),Vs,s(N1),N2):-
  inhabited_type(Xs,[X|Vs],N1,N2).  
inhabited_type(Xs,Vs,s(N1),N3):-
  inhabited_type((X->Xs),Vs,N1,N2),
  inhabited_type(X,Vs,N2,N3).
\end{code}

Clearly the multiset of generated types
has the same count as the set of their
inhabitants and this brings us an
additional 1.5x speed-up.
\begin{code}
inhabited_type(S,T):-inhabited_type(T,[],S,0).
\end{code}

One more (easy) step, giving a {\tt 3x} speed-up,
 makes reaching counts for sizes {\tt 13} and
{\tt 14} achievable: using a faster Prolog,
with a similar {\tt unify\_with\_occurs\_check} built-in,
like  YAP \cite{yap}, with the last value computed
in less than a day.
\BX
The sequence {\bf A220471} completed up to N=14 
\begin{codex}

first 10: 1,2,9,40,238,1564,11807,98529,904318,9006364

11:      96,709,332
12:   1,110,858,977

13:  13,581,942,434
14: 175,844,515,544
\end{codex}
\EX

\section{Doing it once more: generating closed simply-typed normal forms} \label{lnormal}

We will devise similar methods for an important subclass of simply-typed
lambda terms.

\subsection{Generating normal forms}
Normal forms are lambda terms that cannot be further reduced.
A normal form should not be an application with a lambda as its left branch 
and, recursively, its subterms should also be normal forms.
The predicate {\tt normal\_form/2} uses {\tt normal\_form/4} to 
define them  inductively and generates
all normal forms with {\tt S} internal nodes.
\begin{code}
normal_form(S,T):-normal_form(T,[],S,0).

normal_form(v(X),Vs)-->{member(X,Vs)}.
normal_form(l(X,A),Vs)-->down,normal_form(A,[X|Vs]).
normal_form(a(v(X),B),Vs)-->down,normal_form(v(X),Vs),normal_form(B,Vs).  
normal_form(a(a(X,Y),B),Vs)-->down,normal_form(a(X,Y),Vs),normal_form(B,Vs). 
\end{code}

\BX
illustrates closed normal forms with  {\tt 2} internal nodes.
\begin{codex}
?- normal_form(s(s(0)),NF).
NF = l(A, l(B, v(B))) ;
NF = l(A, l(B, v(A))) ;
NF = l(A, a(v(A), v(A))) .
\end{codex}
\EX
The number of solutions of our 
generator replicates entry {\tt A224345} 
in \cite{intseq} that counts closed normal forms of various
sizes.

The predicate {\tt nf\_with\_type} applies the type inference
algorithm to the generated normal forms of size {\tt S}.

\begin{code}
nf_with_type(S,X,T):-normal_form(S,XT),infer_type(XT,X,T).
\end{code}

\subsection{Merging in type inference}

Like in the case of the set of simply-typed lambda terms,
we can define the more efficient
combined generator and type inferrer
predicate {\tt typed\_nf/2}.

\begin{code}
typed_nf(S,X,T):-typed_nf(_XT,X,T,[],S,0).
\end{code}

It works by calling the DCG-expended {\tt typed\_nf/4}
predicate, with the last two arguments enforcing
the size constraints.
\begin{code}
typed_nf(v(V:T),v(V),T,Vs)--> {
   member(V:T0,Vs),
   unify_with_occurs_check(T0,T)
  }.
typed_nf(l(X:TX,A),l(X,NewA),(TX->TY),Vs)-->down,
  typed_nf(A,NewA,TY,[X:TX|Vs]).   
typed_nf(a(v(A),B),a(NewA,NewB),TY,Vs)-->down,
  typed_nf(v(A),NewA,(TX->TY),Vs),
  typed_nf(B,NewB,TX,Vs).
typed_nf(a(a(A1,A2),B),a(NewA,NewB),TY,Vs)-->down,
  typed_nf(a(A1,A2),NewA,(TX->TY),Vs),
  typed_nf(B,NewB,TX,Vs).
\end{code}

\BX
Simply-typed normal forms up to size 3.
\begin{codex}
?- typed_nf(s(s(s(0))),Term,Type).
Term = l(A, l(B, l(C, v(C)))),
Type =  (D->E->F->F) ;
...
Term = l(A, a(v(A), l(B, v(B)))),
Type =  (((C->C)->D)->D) .
\end{codex}
\EX

We are now able to efficiently generate counts for simply-typed normal forms of a given size.

\BX
Counts for closed simply-typed normal forms up to {\tt N=14}.
\begin{codex}
first 10: 1,2,6,23,108,618,4092,30413,252590,2297954

11:     22,640,259
12:    240,084,189
13:  2,721,455,329
14: 32,783,910,297
\end{codex}
\EX
Note that if we would want to just collect the set of types having
inhabitants of a given size, the {\em preservation of typability under
$\beta$-reduction} property \cite{bar93} would
allow us to work with the (smaller) set of simply-typed terms in normal form.

\section{Experimental data} \label{perf}

Figure {\bf 1} gives the number of logical inferences as counted by SWI-Prolog.
This is a good measure of computational effort except for counting operations
like {\tt unify\_{with\_{occurs\_{check}}}} as a single step, while its actual
complexity depends on the size of the terms involved. Therefore,
figure {\bf 2} gives actual timings for the same operations above
{\tt N=5}, where they start to be meaningful.

The ``{\tt closed $\lambda$-terms}'' column gives logical inferences and
timing for generating all closed lambda terms of size given in column 1.
The column ``{\tt gen, then infer}'' covers the algorithm
that first generates lambda terms and then infers their types.
The column ``{\tt gen + infer}'' gives performance data
for the significantly faster algorithm that
merges generation and type inference in the same predicate.
The column ``{\tt inhabitants}'' gives data for the
case when actual inhabitants are omitted in the
merged generation and type inference process.
The column ``{\tt typed normal form}'' shows
results for the fast, merged generation
and type inference for terms in normal form.

As moving from a size to the next typically adds one order of magnitude of computational effort,
computing values for {\tt N=15} and {\tt N=16} is reachable with our best algorithms
for both simply typed terms and their normal form subset.

\begin{figure}
\begin{center}
\begin{tabular}{|r||r|r|r|r|r||}
\cline{1-6}
\cline{1-6}
  \multicolumn{1}{|c||}{{Size}} & 
    \multicolumn{1}{c|}{{ {\small closed $\lambda$-terms}}} &
  \multicolumn{1}{c|}{{ {\small gen, then infer}}} &
  \multicolumn{1}{c|}{{ {\small gen + infer}}} &
  \multicolumn{1}{c|}{{ {\small  inhabitants}}} &
  \multicolumn{1}{c||}{{ {\small typed normal form}}} \\
\cline{1-6} 
\cline{1-6}

1  & 15 & 19 &  16 & 9 & 19      \\ \cline{1-6}
2  & 44 & 59 &  50 & 28 & 47        \\ \cline{1-6}
3  & 166 & 261 &  188 & 113 & 127        \\ \cline{1-6}
4  & 810 & 1,517 &  864 & 553 & 429     \\ \cline{1-6}
5  & 4,905 & 10,930 &  4,652 & 3,112 & 1,814   \\ \cline{1-6}
6  & 35,372 & 92,661 &  28,878 & 19,955 & 9,247    \\ \cline{1-6}
7  & 294,697 & 895,154 &  202,526 & 143,431 & 55,219 \\ \cline{1-6}
8  & 2,776,174 & 9,647,495 &  1,586,880 & 1,146,116 & 377,745 \\ \cline{1-6}
9  & 29,103,799 & 114,273,833 &  13,722,618  & 10,073,400 & 2,896,982 \\ \cline{1-6}
10 & 335,379,436 & 1,471,373,474 &  129,817,948 & 96,626,916 & 24,556,921 \\ \cline{1-6}

\end{tabular} \\
\medskip
\caption{Number of logical inferences used by our generators, as counted by SWI-Prolog 
\label{per1}}
\end{center}
\end{figure}

\begin{figure}
\begin{center}
\begin{tabular}{|r||r|r|r|c|c||}
\cline{1-6}
\cline{1-6}
  \multicolumn{1}{|c||}{{Size}} & 
    \multicolumn{1}{c|}{{ {\small closed $\lambda$-terms}}} &
  \multicolumn{1}{c|}{{ {\small gen, then infer}}} &
  \multicolumn{1}{c|}{{ {\small gen + infer}}} &
  \multicolumn{1}{c|}{{ {\small  inhabitants}}} &
  \multicolumn{1}{c||}{{ {\small typed normal form}}} \\
\cline{1-6} 
\cline{1-6}
5  & 0.001 & 0.001 &  0.001 & 0.000 & 0.001   \\ \cline{1-6}
6  & 0.005 & 0.011 &  0.004 & 0.002 & 0.004   \\ \cline{1-6}
7  & 0.028 & 0.114 &  0.029 & 0.018 & 0.011 \\ \cline{1-6}
8  & 0.257 & 1.253 &  0.242 & 0.149 & 0.050 \\ \cline{1-6}
9  & 2.763 & 15.256 &  2.080 & 1.298 & 0.379 \\ \cline{1-6}
10  & 32.239 & 199.188 &  19.888 & 12.664 & 3.329 \\ \cline{1-6}

\end{tabular} \\
\medskip
\caption{Timings (in seconds) for our generators up to size 10 (on a 2015 MacBook, with 1.3 GHz Intel Core M processor)
\label{perf2}}
\end{center}
\end{figure}

\section{Discussion}\label{disc}
An interesting open problem is if this can be pushed significantly farther. 
We have looked into {\tt term\_hash} based indexing and tabling-based
dynamic programming algorithms, using de Bruijn terms. Unfortunately
as subterms of closed terms are not necessarily closed, even if
de Bruijn terms can be used as ground keys, their associated
types are incomplete and dependent on the context in which
they are inferred.

While it only offers a constant factor speed-up,
parallel execution is a more promising possibility.
However, given the small granularity of the generation
and type inference process, the most useful parallel
execution mechanism would simply split
the task of combined generation and inference process
into a number of
disjoint sets, corresponding to the number
of available processors. A way to do this,
is by using unranking functions (bijections originating in $\N$)
to the sets of combinatorial objects involved, and then,
for $k$ processors, assign work on successive numbers
belonging to the same equivalence class modulo $k$.
Another way is to first generate Motzkin trees and then
launch threads to ``flesh them up'' with logic variables,
run the type inference steps and collect success counts atomically.

We have not seen any obvious way to improve these results
using constraint programing systems, partly because
the   ``inner loop'' computation is unification
with occurs check with computations ranging over
Prolog terms rather than being objects of a constraint domain.
On the other hand, for a given size, exploring grounding to propositional
formulas or answer-set programming seems worth exploring
as a way to take advantage of today's fast SAT-solvers.

Our techniques can be easily adapted to a different size definition like
the ones in \cite{binlamb,maciej16} where variables in de Bruijn notation 
have a size proportional to the distance to their binder.
We have not discussed here the use of similar techniques to improve
the Boltzmann samplers described to \cite{les14bolz}, but clearly
interleaving type checking with the probability-driven
building of the terms would improve their 
performance, by excluding terms with ill-typed subterms
as early as possible, during the large number of retries
needed to overcome the asymptotically 0-density of simply-typed
terms in the set of closed terms \cite{ranlamb09}.

\section{Related work}\label{rels}

The classic reference for lambda calculus is \cite{bar84}.
Various instances of typed lambda calculi are
overviewed in \cite{bar93}.

The combinatorics and asymptotic behavior of various
classes of lambda terms are extensively studied in 
\cite{grygielGen,normalizing13}.
Distribution and density properties
of random lambda terms are described in \cite{ranlamb09}.

Generation of random simply-typed
lambda terms and its applications
to generating
functional  programs from type definitions
 is covered in \cite{fetscher15}.

Several concepts of size have been used in the
literature, partly to facilitate convergence
of formal series in analytic combinatorics
\cite{binlamb,maciej16}.

Asymptotic density properties of simple types (corresponding
to tautologies in minimal logic) have been studied in
\cite{tautintclass} with the surprising
result that ``almost all'' classical tautologies are also 
intuitionistic ones.

In \cite{palka11}  a ``type-directed'' mechanism for
the generation of random terms is introduced,
resulting in more realistic (while not uniformly random) terms,
used successfully in discovering some  bugs in the Glasgow Haskell Compiler (GHC).
A statistical exploration of the structure of the
simple types of lambda terms of a given size
in \cite{iclp15} gives indications that some types
frequent in human-written programs are
among the most frequently inferred ones.

Generators for closed simply-typed 
lambda terms, as well as their normal forms,
expressed as functional programming algorithms,
are given in \cite{grygielGen}, derived from
combinatorial recurrences. However, they
are significantly more complex than the ones
described here in Prolog, and limited
to terms up to size 10.

In the unpublished draft 
\cite{arxiv_play15} we have collected
several lambda term generation
algorithms  written in Prolog and
covering mostly de Bruijn terms
and a compressed de Bruijn representation.
Among them, we have covered linear, affine linear terms
as well as terms of bounded unary height
and in binary lambda calculus encoding.
In \cite{arxiv_play15} type inference algorithms
are also given for SK and Rosser's X-combinator
expressions.
A similar (but 
slower) program for type inference using 
de Bruijn notation is also given in the unpublished draft
\cite{arxiv_play15}, without however
describing the step-by-step derivation
steps leading to it, as done in this paper.

In \cite{fior15} a general constraint logic programming
framework is defined for size-constrained
generation of data structures as well as
a program-transformation mechanism.
While our fine-tuned interleaving of
term generation and type inference directly provides
the benefits of a CLP-based scheme, the program transformation
techniques described in \cite{fior15}  are worth 
exploring for possible performance improvements.

\section{Conclusion} \label{concl}

We have derived several logic programs
that have helped solve the fairly
hard combinatorial counting and generation
problem for simply-typed lambda terms,
4 orders of magnitude higher than
previously published results.

This has put at test two simple but effective
program transformation techniques naturally
available in logic programming languages:
1) interleaving generators and testers by
integrating them in the same predicate and
2) dropping arguments used in generators
when used simply as counters of solutions, as in this case
their role can be kept implicit in the
recursive structure of the program.
Both have turned out to be effective for speeding up computations
without changing the semantics of their 
intended application. We have also managed (after a simple DCG translation)
to work within 
in the minimalist framework of Horn Clauses with sound
unification, showing that non-trivial combinatorics problems
can be handled without any of Prolog's impure features.

Our techniques,  combining 
unification of logic variables
with Prolog's backtracking mechanism
and DCG grammar notation,
recommend logic programming
as a convenient
meta-language for the manipulation
of various families of lambda terms
and the study of their combinatorial
and computational properties.

\section*{Acknowledgement} 
This research has been supported by NSF grant \verb~1423324~. 

We thank the anonymous reviewers of LOPSTR'16 for their constructive 
suggestions and 
the participants of the 9th Workshop
Computational Logic and Applications (\url{https://cla.tcs.uj.edu.pl/})
for enlightening discussions and for
sharing various techniques clarifying the challenges one faces
when having a fresh look at the emerging, interdisciplinary field
of the combinatorics of lambda terms and their applications.

\bibliographystyle{splncs}
\bibliography{theory,tarau,proglang}

\begin{thebibliography}{10}

\bibitem{bar84}
Barendregt, H.P.:
\newblock The Lambda Calculus Its Syntax and Semantics. Revised edn. Volume
  103.
\newblock North Holland (1984)

\bibitem{hindley2008lambda}
Hindley, J.R., Seldin, J.P.:
\newblock Lambda-calculus and combinators: an introduction. Volume~13.
\newblock Cambridge University Press Cambridge (2008)

\bibitem{bar93}
Barendregt, H.P.:
\newblock Lambda calculi with types.
\newblock In: Handbook of Logic in Computer Science. Volume~2.
\newblock Oxford University Press (1991)

\bibitem{palka11}
Palka, M.H., Claessen, K., Russo, A., Hughes, J.:
\newblock Testing an optimising compiler by generating random lambda terms.
\newblock In: Proceedings of the 6th International Workshop on Automation of
  Software Test. AST'11, New York, NY, USA, ACM (2011)  91--97

\bibitem{grygielGen}
Grygiel, K., Lescanne, P.:
\newblock Counting and generating lambda terms.
\newblock J. Funct. Program. \textbf{23}(5) (2013)  594--628

\bibitem{ranlamb09}
David, R., Raffalli, C., Theyssier, G., Grygiel, K., Kozik, J., Zaionc, M.:
\newblock Some properties of random lambda terms.
\newblock Logical Methods in Computer Science \textbf{9}(1) (2009)

\bibitem{bodini11}
Bodini, O., Gardy, D., Gittenberger, B.:
\newblock Lambda-terms of bounded unary height.
\newblock In: ANALCO, SIAM (2011)  23--32

\bibitem{normalizing13}
David, R., Grygiel, K., Kozik, J., Raffalli, C., Theyssier, G., Zaionc, M.:
\newblock Asymptotically almost all $\lambda$-terms are strongly normalizing.
\newblock Preprint: arXiv: math. LO/0903.5505 v3 (2010)

\bibitem{flajolet09}
Flajolet, P., Sedgewick, R.:
\newblock {Analytic Combinatorics}, year = {2009}, isbn = {0521898064,
  9780521898065}, edition = {1}, publisher = {Cambridge University Press},
  address = {New York, NY, USA},

\bibitem{intseq}
Sloane, N.J.A.:
\newblock {The On-Line Encyclopedia of Integer Sequences.}
\newblock (2014) ~Published electronically at https://oeis.org/.

\bibitem{padl15}
Tarau, P.:
\newblock {On Logic Programming Representations of Lambda Terms: de Bruijn
  Indices, Compression, Type Inference, Combinatorial Generation,
  Normalization}.
\newblock In Pontelli, E., Son, T.C., eds.: {Proceedings of the Seventeenth
  International Symposium on Practical Aspects of Declarative Languages
  PADL'15}, Portland, Oregon, USA, Springer, LNCS 8131 (June 2015)  115--131

\bibitem{cicm15}
Tarau, P.:
\newblock {Ranking/Unranking of Lambda Terms with Compressed de Bruijn
  Indices}.
\newblock In Kerber, M., Carette, J., Kaliszyk, C., Rabe, F., Sorge, V., eds.:
  {Proceedings of the 8th Conference on Intelligent Computer Mathematics},
  Washington, D.C., USA, Springer, LNAI 9150 (July 2015)  118--133

\bibitem{ppdp15tarau}
Tarau, P.:
\newblock { On a Uniform Representation of Combinators, Arithmetic, Lambda
  Terms and Types}.
\newblock In Albert, E., ed.: {PPDP'15: Proceedings of the 17th international
  ACM SIGPLAN Symposium on Principles and Practice of Declarative Programming},
  New York, NY, USA, ACM (July 2015)  244--255

\bibitem{iclp15}
Tarau, P.:
\newblock {On Type-directed Generation of Lambda Terms}.
\newblock In De~Vos, M., Eiter, T., Lierler, Y., Toni, F., eds.: {31st
  International Conference on Logic Programming (ICLP 2015), Technical
  Communications}, Cork, Ireland, CEUR (September 2015) available online at
  http://ceur-ws.org/Vol-1433/.

\bibitem{arxiv_play15}
Tarau, P.:
\newblock A logic programming playground for lambda terms, combinators, types
  and tree-based arithmetic computations.
\newblock CoRR \textbf{abs/1507.06944} (2015)

\bibitem{StanleyEC}
Stanley, R.P.:
\newblock Enumerative Combinatorics.
\newblock Wadsworth Publ. Co., Belmont, CA, USA (1986)

\bibitem{statman79}
Statman, R.:
\newblock {Intuitionistic Propositional Logic is Polynomial-Space Complete}.
\newblock Theor. Comput. Sci. \textbf{9} (1979)  67--72

\bibitem{swi}
Wielemaker, J., Schrijvers, T., Triska, M., Lager, T.:
\newblock {SWI-Prolog}.
\newblock Theory and Practice of Logic Programming \textbf{12} (1 2012)  67--96

\bibitem{yap}
Costa, V.S., Rocha, R., Damas, L.:
\newblock {The YAP Prolog system}.
\newblock Theory and Practice of Logic Programming \textbf{12} (1 2012)  5--34

\bibitem{binlamb}
Grygiel, K., Lescanne, P.:
\newblock {Counting and Generating Terms in the Binary Lambda Calculus
  (Extended version)}.
\newblock CoRR \textbf{abs/1511.05334} (2015)

\bibitem{maciej16}
Bendkowski, M., Grygiel, K., Lescanne, P., Zaionc, M.:
\newblock {A Natural Counting of Lambda Terms}.
\newblock In Freivalds, R.M., Engels, G., Catania, B., eds.: {SOFSEM} 2016:
  Theory and Practice of Computer Science - 42nd International Conference on
  Current Trends in Theory and Practice of Computer Science, Harrachov, Czech
  Republic, January 23-28, 2016, Proceedings. Volume 9587 of Lecture Notes in
  Computer Science., Springer (2016)  183--194

\bibitem{les14bolz}
Lescanne, P.:
\newblock Boltzmann samplers for random generation of lambda terms.
\newblock CoRR \textbf{abs/1404.3875} (2014)

\bibitem{fetscher15}
Fetscher, B., Claessen, K., Palka, M.H., Hughes, J., Findler, R.B.:
\newblock Making random judgments: Automatically generating well-typed terms
  from the definition of a type-system.
\newblock In: Programming Languages and Systems - 24th European Symposium on
  Programming, {ESOP} 2015, Held as Part of the European Joint Conferences on
  Theory and Practice of Software, {ETAPS} 2015, London, UK, April 11-18, 2015.
  Proceedings. (2015)  383--405

\bibitem{tautintclass}
Genitrini, A., Kozik, J., Zaionc, M.:
\newblock {Intuitionistic vs. Classical Tautologies, Quantitative Comparison}.
\newblock In Miculan, M., Scagnetto, I., Honsell, F., eds.: Types for Proofs
  and Programs, International Conference, {TYPES} 2007, Cividale del Friuli,
  Italy, May 2-5, 2007, Revised Selected Papers. Volume 4941 of Lecture Notes
  in Computer Science., Springer (2007)  100--109

\bibitem{fior15}
Fioravanti, F., Proietti, M., Senni, V.:
\newblock Efficient generation of test data structures using constraint logic
  programming and program transformation.
\newblock Journal of Logic and Computation \textbf{25}(6) (2015)  1263--1283

\end{thebibliography}

\begin{codeh}
genDBterm(L,T):-genDBterm(T,0,L,0).

genDBterm(v(X),V)-->
  {up_to(V,s(X))}.
genDBterm(l(A),V)-->down,
  {down(NewV,V)},
  genDBterm(A,NewV).
genDBterm(a(A,B),V)-->down,
  genDBterm(A,V),
  genDBterm(B,V).  

up_to(X,X).
up_to(s(X),R):-up_to(X,R).

genTypedDBTerm(S,X,T):-
  genTypedDBTerm(X,T,[],S,0).

genTypedDBTerm(v(I),V,Vs)--> {
   select_nth(I,Vs,V0),
   unify_with_occurs_check(V,V0)
  }.
genTypedDBTerm(a(A,B),Y,Vs)-->down,
  genTypedDBTerm(A,(X>Y),Vs),
  genTypedDBTerm(B,X,Vs).
genTypedDBTerm(l(A),(X>Y),Vs)-->down,
  genTypedDBTerm(A,Y,[X|Vs]).

select_nth(0,[X|_],X).
select_nth(s(I),[_|Xs],Y):-select_nth(I,Xs,Y).
     

counts(M,Goal):-
  functor(Goal,F,_),writeln(F:M),
  findall(T,(between(1,M,N),n2s(N,S),arg(1,Goal,S),sols(Goal,T),writeln(N:T)),Ts),
  writeln(counts=Ts),
  ratios(Ts,Rs),
  writeln(ratios=Rs).

sols(Goal, Times) :-
        Counter = counter(0),
        (   Goal,
            arg(1, Counter, N0),
            N is N0 + 1,
            nb_setarg(1, Counter, N),
            fail
        ;   arg(1, Counter, Times)
        ).
  
times(M,Goal):-
  functor(Goal,F,_),writeln(F:M),
  between(1,M,N),
  n2s(N,S),arg(1,Goal,S),
  writeln(N:F),
  time((Goal,fail;true)),
  fail.

ratios([X|Xs],Rs):-
  map_ratio(Xs,[X|Xs],Rs).

map_ratio([],[_],[]).
map_ratio([X|Xs],[Y|Ys],[R|Rs]):-
  R is X/Y,
  map_ratio(Xs,Ys,Rs).

show(N,Goal):-
  functor(Goal,F,_),
  writeln(F:N),
  n2s(N,S),
  arg(1,Goal,S),
    Goal,
    show_one(Goal),
  fail.

show_one(Goal):-
  numbervars(Goal,0,_),
  Goal=..[_,_|Xs],
    member(X,Xs),
    writeln(X),
  fail
; nl.

cgo:-
  N=6,Gs=[cm,cl,cp,ct,clt,ctl,cit,cnf,cnt],  
  member(G,Gs),
  call(G,N),
  fail.
  

cm(N):- counts(N,motzkin(_,_)).            
cl(N):-counts(N,lambda(_,_)).                
cp(N):-counts(N,partitions(_,_)).          
ct(N):-counts(N,maybe_type(_,_,_)).        
clt(N):-counts(N,lamb_with_type(_,_,_)).   
ctl(N):-counts(N,typed_lambda(_,_,_)).       
cit(N):-counts(N,inhabited_type(_,_)).     
cnf(N):-counts(N,normal_form(_,_)).        
cnt(N):-counts(N,nf_with_type(_,_,_)).     
ctn(N):-counts(N,typed_nf(_,_,_)).         

cdb(N):-counts(N,genTypedDBTerm(_,_,_)).


bgo:-bgo(7).

bgo(N):-
  Gs=[bm,bl,bp,bt,blt,btl,bit,bnf,bnt],  
  member(G,Gs),
  call(G,N),
  fail.

bgo1:-N=10,
  Gs=[bl,blt,btl,bit,btn],  
  member(G,Gs),
  call(G,N),
  fail.  
  
bm(N):-times(N,motzkin(_,_)).
bl(N):-times(N,lambda(_,_)).
bp(N):-times(N,partitions(_,_)).
bt(N):-times(N,maybe_type(_,_,_)).
blt(N):-times(N,lamb_with_type(_,_,_)).
btl(N):-times(N,typed_lambda(_,_,_)).
bit(N):-times(N,inhabited_type(_,_)).
bnf(N):-times(N,normal_form(_,_)).
bnt(N):-times(N,nf_with_type(_,_,_)).
btn(N):-times(N,typed_nf(_,_,_)).

bdb(N):-times(N,genTypedDBTerm(_,_,_)).
  
  
sgo:-
  N=2,Gs=[sm,sl,sp,st,slt,stl,sit,snf,snt],  
  member(G,Gs),
  call(G,N),
  fail.  
    
sm(N):-show(N,motzkin(_,_)).
sl(N):-show(N,lambda(_,_)).
sp(N):-show(N,partitions(_,_)).
st(N):-show(N,maybe_type(_,_,_)).
slt(N):-show(N,lamb_with_type(_,_,_)).
stl(N):-show(N,typed_lambda(_,_,_)).
sit(N):-show(N,inhabited_type(_,_)).
snf(N):-show(N,normal_form(_,_)).
snt(N):-show(N,nf_with_type(_,_,_)).
stn(N):-show(N,typed_nf(_,_,_)).

sdb(N):-show(N,genTypedDBTerm(_,_,_)).    
    

texshow(T):-
  numbervars(T,0,_),
  texshow(T,Cs,[]),
  maplist(write,Cs),
  fail.
texshow(_).

texshow(v('$VAR'(I)))--> [x],['_'],[I].
texshow(l(('$VAR'(I)),E))-->[(' ')],[('~\\lambda ')],[x],['_'], [I],[('.')],texshow(E),[(' ')].
texshow(a(X,Y))-->[('(')],texshow(X),[('~')],texshow(Y),[(')')].


nv(X):-numbervars(X,0,_).

pp(X):-nv(X),writeln(X),fail;true.
       

scomb(l(A, l(B, l(C, a(a(v(A), v(C)), a(v(B), v(C))))))).    
  
ycomb(l(A, a(l(B, a(v(A), a(v(B), v(B)))), l(C, a(v(A), a(v(C), v(C))))))).
   
\end{codeh}

\end{document}